\documentclass[final]{svjour2}

\usepackage{graphicx}
\usepackage{rotating}
\usepackage{amssymb}
\usepackage{mathptmx}
\usepackage[numbers]{natbib}
\usepackage[T1]{fontenc}
\usepackage[latin1]{inputenc}

\bibpunct{}{}{,}{s}{}{,}

\begin{document}

\title{Electron- nuclear recoil discrimination by pulse shape analysis.}

\author{J.~Elbs$^1$ \and Yu.~M.~Bunkov$^1$ \and E.~Collin$^1$ \and H.~Godfrin$^1$ \and O.~Suvorova$^2$}

\institute{1: Institut Néel, CNRS/UJF, 25, av. des Martyrs, 38042 Grenoble, France\\ Tel.: +33 4 76 88 12 52\\ Fax: +33 4 76 87 50 60\\
\email{yuriy.bunkov@grenoble.cnrs.fr}\\
2: Institute for Nuclear Research, RAS, 117312, Moscow, Russia
}

\date{\today}

\maketitle


\begin{abstract}

In the framework of the ``ULTIMA'' project, we use ultra cold
superfluid $^3\!$He bolometers for the direct detection of single
particle events, aimed for a future use as a dark matter detector.
One parameter of the pulse shape observed after such an event is the
thermalization time constant $\tau_b$. Until now it was believed
that this parameter only depends on geometrical factors and
superfluid $^3\!$He properties, and that it is independent of the
nature of the incident particles. In this report we show new results
which demonstrate that a difference for muon- and neutron events, as
well as events simulated by heater pulses exist. The possibility to
use this difference for event discrimination in a future dark matter
detector will be discussed.
\end{abstract}
\section{INTRODUCTION}

$^3\!$He is a very appealing target matter for the direct search of
non-baryonic dark matter\cite{Pickett88, Bunkov95}. Various
attractive features special to $^3\!$He can be pointed out: the
large neutron capture cross section provides an inherent
discrimination mechanism \linebreak
against neutrons, the particle which is normally hardest to reject.
The large number of unpaired nucleons per unity of mass gives an advantage in a large number
of WIMP models which predict a spin dependent interaction\cite{Mayet1, Mayet2}. This
sensitivity to the so-called axial channel makes $^3\!$He complementary to the most advanced
detectors like Edelweiss\cite{Edelweiss} and CDMS\cite{CDMS}, which are mainly sensitive to
the spin independent interaction. The $^3\!$He atoms being much lighter than the expected
WIMP masses, the recoil energy would be limited to $E<6$~keV, allowing for an additional
discrimination by rejection of events above this threshold, and thus an enhanced  signal
to noise ratio. The absolute purity of $^3\!$He at the working temperatures of $\approx$130~$\mu$K
prevents the presence of false event producing radioactive impurities in the sensitive medium itself.
The large transparency to $\gamma$-rays reduces the expected false event rate further. Because of the
existence of only one thermal bath, the Bogoliubov quasiparticle excitations, problems of two or more
badly coupled thermal baths do not arise\cite{twobaths}.\\
Numerical simulations show that with a large bolometer, a
sensitivity to neutralinos within a large number of supersymetric
models would be obtained\cite{Mayet1, Mayet2}. Additionally those
simulations show that using a matrix design of many bolometric cells
should provide a very good overall rejection to background events.
Using a 3-cell prototype, a certain number of the requirements of a
future dark matter detector have already been demonstrated: A 1~keV
sensitivity has been achieved\cite{Winkel07}, with further potential
of improvement by replacement of the currently used NbTi Vibrating
Wire Resonator thermometers with specially designed, microfabricated
Silicon resonators. Coincidences between neighboring cells, mainly
produced by cosmic muon events, were observed and support the idea
of a discrimination by coincident measurement. The possibility of a
precise energy calibration by introduction of a well measured amount
of energy using a second Vibrating Wire Resonator was first
described in Ref.\cite{Bauerle98}, and many details can be found in
Ref.\cite{Winkel07} . This calibration is special as it does not
rely on the use of a radioactive reference source as for most other
bolometric detectors.\\
For a future dark matter detector, an alternative channel of
discrimination would be highly desirable. The main idea is the
parallel measurement of either scintillation light or ionization
charges. The ratio of energy going to scintillation/ionisation is
observed to be different between particles interacting by electronic
and nuclear recoil in the case of $^4\!$He\cite{McKinsey}, and the
differences found between heat measured and total heat release
expected in superfluid $^3\!$He bolometers show that this effect
persists in our working conditions\cite{Winkel07}. An experiment
testing the realization of the ionization channel is in preparation.
In this paper we show experimental result which indicate that a
simple pulse shape analysis could provide an additional
discrimination mechanism.

\section{EVENT SHAPE}
The heart of our particle detectors are small cylindrical copper
cells of Volume $V=$0.13~cm$^3$ filled with superfluid $^3\!$He-B,
similar to one, proposed by Fisher {\it et al.}\cite{Fisher92}.

\begin{figure}
    \centering
        \includegraphics[width=0.60\textwidth]{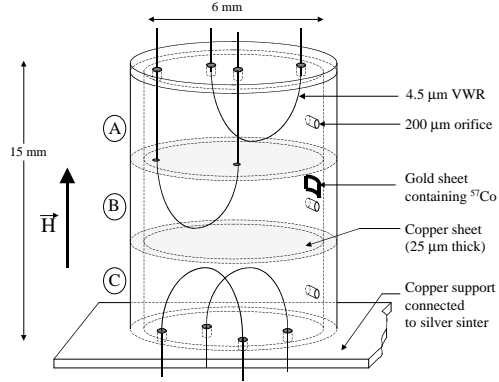}
        \caption{Schematic view of the 3 cell bolometer. The cells are filled with superfluid
        $^3\!$He and immersed into a superfluid $^3\!$He bath, which functions as the heat sink.
        The thermal contact is obtained by the small orifices. Each cell contains a VWR thermometer
        and can be operated independently. One of the cell contains a second VWR for doing calibration pulses.
        The $^{57}\!$Co source in another cell was used to demonstrate the 1~keV sensitivity\cite{Winkel07}.  }
    \label{fig:3cells}
\end{figure}

These cells are immersed inside a superfluid $^3\!$He-B bath.
Thermal contact between the cells and the bath is achieved by small
orifices of $S=200$~$\mu$m diameter. When a particle interacts
inside the cell, the corresponding energy deposition will heat up
the cell, and the quasiparticle density will rise. Subsequently, the
cell will thermalize with the outer bath with a time constant
$\tau_b$ on the order of a few seconds. The main idea is to
demonstrate experimentally that the discrimination mechanism
proposed\cite{Mayet1} is feasible: strongly interacting particle, in
our case mainly cosmic muons, which cross several cells will leave
simultaneously an energy in all the cells along their track. On the
other hand, weakly interacting particles will leave an energy in
only one cell. A scheme of the 3 cell bolometer is presented in
fig.~\ref{fig:3cells}.

An important part for the functioning of the bolometers is the thermometry. High
precision thermometry is achieved by the use of Vibrating Wire Resonators (VWR)\cite{Guenault86, Winkel04}.
The VWR we use is a fine superconducting NbTi wire, bent into a semi circular shape, excited by the Laplace
force. When driven well below the pair breaking velocity, the interaction of a VWR with the
superfluid $^3\!$He-B is given by the scattering of quasiparticles on the wire surface, and a frictional
force linear with velocity is observed. A frequency sweep over the VWR resonance thus delivers
a Lorentzian lineshape. The Full Width at Half Maximum (FWHM) $W(T)$ depends on the quasiparticle (QP)
density which changes exponentially with temperature. In the zero field limit, $W(T)$ writes\cite{Fisher91}
\begin{equation}
\label{eqWidth}
W(T)=\alpha \exp (-\Delta/k_B T),
\end{equation}
where $\Delta$ is the superfluid gap at zero temperature and zero field, and $\alpha$ a prefactor
which depends on the geometry of the VWR and on properties of the liquid. It is this relation which
will lead us to use in the following the terms {\it VWR width} and {\it temperature} interchangeably.

After a heating event caused for example by a cosmic particle, the
QP density inside the cell will suddenly rise, and then go back to
the initial QP density by thermalization via the hole. This
relaxation is an exponential process, with the time constant
$\tau_b$ determined by the geometry, i.e. mainly the ratio $V/S$ and
the QP mean group velocity $\bar{v}_g$. We thus can write the
equilibrium VWR width after an event at $t_0$, of amplitude $A$ and
at base temperature $W_0$ as
\begin{equation}
    W_{eq}(t)=W_0 + Ae^{-\frac{t-t_0}{\tau_b}}\Theta(t-t_0),
    \label{eq:WidthEq}
\end{equation}
with $\Theta(t)$ the Heavyside step function.
For a flat hole, i.e. a hole with diameter~$\gg$~thickness, $\tau_b$ can be calculated, considering
that effects introduced by textures are negligible:
\begin{equation}
    \tau_b=4\frac{V}{S}\frac{1}{\bar{v}_g}
    \label{eq:taub}
\end{equation}
As $\bar{v}_g$ shows a $T^{1/2}$ dependence, $\tau_b$ is expected to only vary slowly with temperature.
It must be pointed out that our hole has a diameter of 200~$\mu$m and a thickness of 400~$\mu$m,
which should increase the real value considerably due to the possibility of backscattering in
the hole on rough surfaces, and via Andreev scattering.

\begin{figure}
    \centering
    \includegraphics{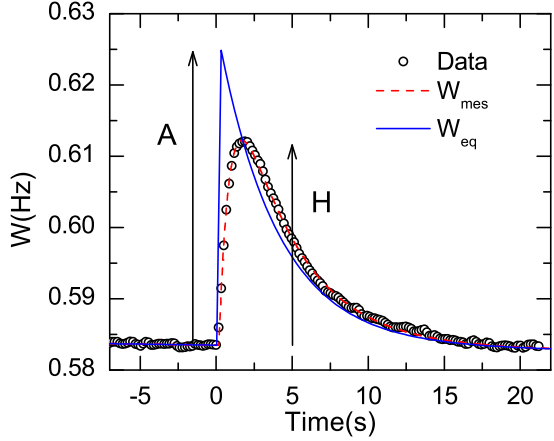}
    \caption{(Color on-line) A typical event, probably caused by a cosmic muon (circles). The data can be fitted (dashed line) with eq.~\ref{eq:WidthRet}. Using the obtained fit parameters $A$ and $\tau_b$, the equilibrium linewidth and thus the quasiparticle density can be illustrated using eq.~\ref{eq:WidthEq} (bold line).}
    \label{PeakandFit}
\end{figure}

The VWR being a high Q macroscopic object, it can not immediately adapt to a new equilibrium
position after a quick temperature change, but does so with a delay, determined by the time
constant $\tau_w=\frac{1}{\pi W_0}$. The measured wire linewidth is thus
\begin{equation}
W_{mes}(t)=W_0 + A(e^{-\frac{t-t_0}{\tau_b}}-e^{-\frac{t-t_0}{\tau_w}})\Theta(t-t_0).
\label{eq:WidthRet}
\end{equation}
A typical particle event together with a fit using this formula is presented in fig.~\ref{PeakandFit}.
As can be seen, this formula seems to represent very well the experimentally obtained data.
One important prerequisite for the above formula to work has not
been mentioned yet: the usual hypothesis is that the internal
thermalization, i.e. the processes which lead from the initial, very
localized, ionization to a homogeneous thermal quasiparticle
distribution inside the cell, is very fast compared to the other
time constants involved. The initial very high energetic
quasiparticles thermalize very fast by creation of new pairs of
quasiparticles. The final stage of thermalization for near thermal
QP is supposed to happen by nonlinear processes on the cell walls,
the corresponding time constant should be on the order of the time
of flight of the quasiparticles towards the cell walls. This time
constant is obtained by dividing the cell dimensions by the mean
group velocity and amounts in our case to $\approx$0.3~ms, which is
three orders of magnitudes smaller than typical VWR response times.
As a result, a sharp slope of the rising edge should be observed
which is in good agreement with experimental data
(fig.~\ref{PeakandFit}). The expectation is thus that internal
thermalization is virtually instantaneous, and that no information
about the nature of the influence survives long enough to have an
influence on our measurements.

\section{PEAK SHAPE ANALYSIS}
\begin{figure}
    \centering
    \includegraphics{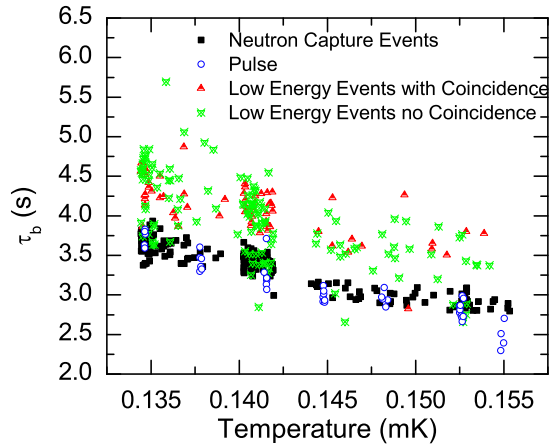}
    \caption{(Color on-line) $\tau_b$ parameter for a large number of particle events and heater pulses as
    a function of temperature. Particle events releasing an energy of about 650~keV (black squares)
    are identified as neutron capture events. Events releasing less than this energy are separated in
    events showing and not showing coincidence with the neighboring cell. Two bands of values of $\tau_b$
    can be clearly identified, with all neutron capture events and heater pulses being in the lower band,
    and all events showing coincidence in the upper band.}
    \label{fig:taubTemp}
\end{figure}

In order to study whether all of the above is confirmed by our
experimental data, a large number of measured peaks, acquired during
about 48~h in the presence of an AmBe neutron source has been fitted
using eq.~\ref{eq:WidthRet}. For the fits, the baseline $W_0$ has
been determined first by fitting the signal before an event with a
constant. Then the main fit with eq.~\ref{eq:WidthRet} has then been
done with the three free parameters $A$, $\tau_b$, $t_0$. As for the
moment we were mainly interested in obtaining clear results for the
$\tau_b$ parameter, the analysis is limited to relatively high
energetic events with $E$>100~keV. An improved analysis which
potentially extends the analysed region down to 25~keV events is a
work in progress. The results as a function of temperature are
presented in fig.~\ref{fig:taubTemp}. In this figure, the particle
events can be separated by two discrimination mechanisms in three
groups: Firstly, it is well established that neutrons have a high
cross section to undergo a neutron capture reaction. This releases a
well defined energy of 764~keV, of which about 655~keV are released
as heat. The remaining events can be separated into events leaving
or not leaving at the same time an energy in an adjacent cell. The
events showing coincidence are identified as muons. The last group
of events which show no coincidence correspond to either cosmic
muons, crossing on its trajectory none of the other cells, or an
event caused by other particles, most likely the elastic scattering
with neutrons (neutron recoil event).

The first remark which can be made is that  a temperature dependence does indeed exist. A $T^{-1/2}$
law is expected due to the temperature dependence of the mean group velocity, but the observed data fits
better with a $T^{-2}$ law. Additionally, the observed absolute values are for example at 140~$\mu$K more
than 5 times higher than the calculated value of 0.64~s using the flat hole approximation
(eq.~\ref{eq:taub}). The reason for this quite strong temperature dependence might be that textures
in and near the orifice prevent the lowest energetic excitations to leave the cell, as they might be
forced to Andreev backscatter even by small energy barriers.

\begin{figure}
    \centering
    \includegraphics{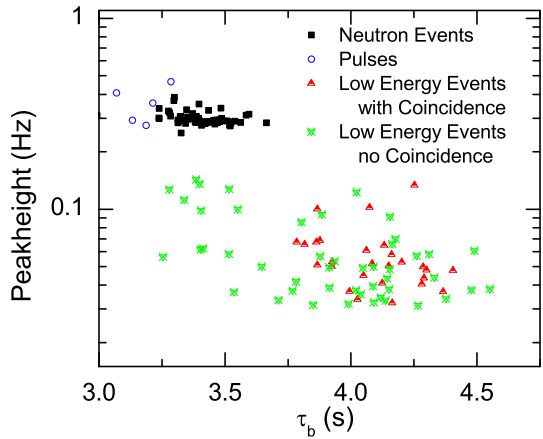}
    \caption{(Color on-line) For a small slice of temperature (0.140~mK$<T<$0.143~mK), the peak height is
    plotted as a function of the $\tau_b$ parameter. The calibration factor as determined by heater
    pulses is $\sigma=$0.44~mHz/keV. The neutron capture events are clearly identified by the narrow
    band of energies around 655~keV (0.29~Hz). For lower energies, we identify the region of high
    values of $\tau_b$ as muon events, and the region of low values of $\tau_b$ as neutron recoil events.
    It must be stressed out that our energy resolution using algorithms which do not try to find good
    values for $\tau_b$ is much better (in the best cases around 1~keV) than what this graph suggests.
    An improved analysis which tries to extent this graph to lower energies is a work in progress.}
    \label{fig:taubAmpl}
\end{figure}

In fig.~\ref{fig:taubAmpl} a part of the data for the temperature slice 0.140~mK$<T<$0.143~mK is plotted
with the energy resolved on the y-axis. The main observation from both graphs is the appearance of two
different bands of values. All neutron capture events and all heater pulses show low values of $\tau_b$,
with the heater pulses in average even smaller than the neutron capture events. All muon events
identifiable by coincident pulses show high values of $\tau_b$. Non coincident low energy events are
mainly found in the upper band, but some are found in the lower band. As explained, muon events can,
but not necessarily have to show coincidence. On the other hand, the neutron source was placed in the
plane horizontal to the cell alignment. This, and the relatively small recoil cross section means that
coincidences for neutron recoil events should be very rare. In the absence of a neutron source, the lower
band completely disappears, these events with low $\tau_b$ are thus clearly linked to the presence of the
neutron source. We thus identify in fig.~\ref{fig:taubAmpl} the region with $E<$600~keV and $\tau_b$>3.7~s
as events caused by cosmic muons, and the region with $E<$600~keV and $\tau_b$<3.6~s as neutron recoil events.

\section{Discussion and application for a future dark matter detector}
It is not easy to imagine what can be the origin of this effect. It
is necessary to find some mechanism which lives long enough to have
an influence on the order of seconds, and which at the same time is
different for different type of particles depositing an energy. For
instance our best guess is a delayed heat release from metastable
triplet $^3\!$He dimers. Scintillation measurements done on
scintillation in $^4\!$He and $^3\!$He show that their radiative
life constant is about 13~s. Additionally it has been shown that
during the rapid processes after the primary ionization a large
amount of these dimers form. The fraction of deposited energy stored
in these metastable state depend on the particle interacting. For
light particles like muons, electrons and $\gamma$ this fraction is
about 25\%, for heavy particles it is only about 7\%. As a result,
the time delayed energy release is for neutrons about three times
smaller than for light particles. First numerical simulations based
on a non radiative deexcitation, and thus a partially delayed heat
release look very promising.

The current work inscribes itself in a long term project (ULTIMA)
which intends to build a dark matter detector with superfluid
$^3\!$He as the target matter. One crucial of every dark matter
detector is an efficient method to distinguish between events caused
by ordinary particles and the weakly interacting particles which are
searched. Numerical simulations using the GEANT4 code\cite{Mayet2}
already showed that a very high overall rejection factor can be
obtained using a high number of bolometric cells arranged in a
matrix. Nevertheless, a second channel of discrimination would be
highly desirable. The possibility to measure ionization at the same
time as the heat release is under investigation. As the ratio heat
release/ionisation is expected to depend on the nature of the
particle interaction, this has good potential to provide a reliable
alternative discrimination, but at the cost of an increased
complexity of the experimental setup.

The interest for the effect presented in this paper is that it does
not come at an additional experimental cost. Nevertheless, in order
to use this pulse shape analysis for discrimination, it must be
shown in future experiments that the increased value of $\tau_b$
exists not only for muons, but in general for particles interacting
by electronic recoil, and that the effect prevails down to the
energy range of interest (1-6~keV) which can be foreseen for
neutralino dark matter. Additionally it is highly desirable to get a
good understanding for the underlying mechanism, in order to
optimize the future bolometers.

And finally, the studies of time delated heat relies from exited
dimers well support the results of experiments of "Big Bang"
simulation in superfluid $^3\!$, published in Ref.\cite{BigBang}. At
that experiments the time constant of cell recovering was much
longer, about a 60 s. and the heat relies from dimers recombination
counted in the balance of energy as a heating. The energy losses for
scintillation, suggested in Ref.\cite{Leggett}, are overestimated on
factor of two or more. Consequently, the conclusions of article
\cite{Leggett} has no experimental ground. The article with proper
consideration of energy distribution for different processes for
neutrons and muons events is under preparation.

\section*{ACKNOWLEDGMENTS}
We thank Prof. A. Parshin, Prof. G. Seidel and Dr. C. Winkelmann for
useful discussions. This work was done in the framework of the
ULTIMA project of the ``Agence Nationale de la Recherche", France
(NTO5-2\_41909) and under a support of French-Russian program of
cooperation (Institute NEEL, CNRS - Institute Kapitza, RAS, project
19058) .

\end{document}